\documentclass[twocolumn]{revtex4-1} %
\usepackage{graphicx,graphicx,epsfig, subfigure}

\usepackage{graphicx}
\usepackage{bm}

\newcommand{\la}{\langle}
\newcommand{\ra}{\rangle}

\newcommand{\rar}{\rightarrow}
\newcommand{\da}{\downarrow}
\newcommand{\ua}{\uparrow}
\newcommand{\be}{\begin{equation}}
\newcommand{\ee}{\end{equation}}

\begin{document}
\title{Nonequilibrium spin noise spectroscopy}

\author{Fuxiang Li}
\affiliation{Center for Nonlinear Studies, Los Alamos National Laboratory,
  Los Alamos, NM 87545 USA}
 \affiliation{Department of Physics, Texas A\&M University, College Station, TX 77845 USA}
\author{Yuriy V. Pershin}
\affiliation{Department of Physics and Astronomy, University of South Carolina, Columbia, SC, 29208 USA}

\author{Valeriy A. Slipko}
\affiliation{Department of Physics and Astronomy, University of South Carolina, Columbia, SC, 29208}
\affiliation{ Department of Physics and Technology, V. N. Karazin
Kharkov National University, Kharkov 61077, Ukraine}




\author{N. A. Sinitsyn}
\email{nsinitsyn@lanl.gov}
\affiliation{Theoretical Division, Los Alamos National Laboratory,
  Los Alamos, NM 87545 USA}


\begin{abstract}
 Spin Noise Spectroscopy  (SNS)  is an experimental approach to obtain correlators of mesoscopic spin fluctuations in time by purely optical means. We explore the information that this technique can provide when it is applied to
  a weakly non-equilibrium regime when an electric current is driven through a sample by an electric field. We find that the noise power spectrum of conducting electrons experiences a shift, which is proportional to the strength of the spin-orbit coupling for electrons moving along the electric field direction. We propose applications of this effect to measurements of spin orbit coupling anisotropy and separation of spin noise of conducting and localized electrons.
\end{abstract}

\date{\today}

\maketitle

{\it Introduction.}
Optical spin noise spectroscopy (SNS) \cite{Aleksandrov81} has been recently introduced as a promising approach for probing local  fluctuations of spins in semiconducting materials \cite{Oestreich05, Muller08, Crooker09, Oestreich10, cond-SNS, Huang11, ultrahigh}, atomic gases \cite{Crooker04, Minhaila06, Minhaila06_2, Katsoprinakis07, Shah10, Zapasskii13} and quantum dots \cite{ WarburtonArxiv13,Crooker10, Li12}.
In particular, if $S_z(t)$ is the time-dependent polarization of spins in a mesoscopic region of a semiconductor, SNS can be used to obtain the spin noise power
spectrum \cite{simple-noise, Kos10, Glazov11, Pershin12, sinitsyn-12prl,roy-13}:
 \be
 P(\omega) =2\int_{0}^{\infty}dt~ \cos {(\omega t)} \la S_z(t)S_z(0)\ra.
\label{SNPS}
\ee
The advantage of the optical SNS over the other measurement approaches (e.g., optical pump-probe \cite{review, crooker-07njp} or STM measurements of a single spin \cite{STM}) is usually associated with the minimal energy dissipated, i.e., it can probe  spin dynamics at thermodynamic equilibrium. In addition, the SNS allows accumulation of a large statistics that smoothes out the statistical noise in the data, so one can study subtle details at the tails of  spin-spin correlators \cite{Li12}. Sensitivity of this approach is continuously improving, e.g. the recently introduced Ultrahigh Bandwidth SNS  can resolve spin correlations with picosecond resolution \cite{ultrahigh}.

At the thermodynamic equilibrium, the fluctuation-dissipation theorem predicts that the knowledge of the spin correlator $\la S_z(t)S_z(0)\ra$, which is obtained by SNS, is formally equivalent to the information that can be obtained from a linear response pump-probe measurements. So, for systems at the thermodynamic equilibrium, SNS probes characteristics that, at least in principle, can be obtained from more traditional approaches based on characterizing system's linear response.

In this letter we explore the possibility to apply SNS  to semiconductors in a non-equilibrium steady state. The behavior of the spin-spin correlator in a non-equilibrium regime may no longer be the subject of the  fluctuation-dissipation theorem. Hence, even if a perturbation from the equilibrium is weak (in our case it will be a weak electric field that induces an electric current), identifying non-equilibrium contribution to the spin correlator may provide the information about the system that cannot be obtained from the linear response characteristics.

For a demonstration, we consider  the effect of an electric field on spin fluctuations of conducting elections in 2D electron gas with Rashba and Dresselhaus spin orbit couplings and spin-independent scatterings. We develop an approach that allows us not only to derive the equations for mean spin polarization \cite{bleibaum} but also to relate parameters of spin fluctuations to the shot noise at microscopic scattering events.

{\it Stochastic dynamics of spin fluctuations.} Consider spin fluctuations from the mean steady state of an electron system. Let  ${ \hat{\rho}}_{\bf k}$ be the spin density matrix in the momentum space, which is a 2$\times$2 matrix in spin indexes.  
We assume that the observation region is much larger than the spin diffusion length, so that we can consider dynamics only in the momentum  space \cite{Kohn-Luttinger}. The evolution of the spin density matrix in a momentum space volume  ${\bf k}$ is described by the quantum Boltzmann equation:
\begin{equation}
\dot{{\hat{\rho}}}_{\bf k}-e{\bf E}\cdot\nabla_{\bf k}{\hat{\rho}}_{\bf k}+i[\hat{H}_0, { \hat{\rho}_{\bf k}}]=\hat{I}_{coll},
\label{kieq}
\end{equation}
in which $\hat{I}_{coll}$ is the collision term due to elastic scattering on impurities, and $\hat{H}_0$ is the scattering-free part of the Hamiltonian:
\begin{equation}
\hat{H}_0=\frac{k^2}{2m}-\frac{1}{2}{\bm \Omega}_{\bf k}\cdot {\bm \sigma}-\frac{1}{2}{\bf H}\cdot{\bm \sigma},
\end{equation}
where $\bm \sigma$ is the vector of Pauli matrices,  ${\bf H}$
is the in-plane magnetic field (with absorbed Bohr magneton and g-factor). 

The spin orbit coupling field can be written as a sum of two parts ${\bm \Omega}_{\bf k}=k_x{\bm \Omega}_1+k_y{\bm \Omega}_2$, with:
\begin{equation}
{\bm \Omega}_1=2(\beta\hat{x}-\alpha\hat{y}),\, \, \, {\bm \Omega}_2= 2(\alpha\hat{x}-\beta\hat{y}),
\label{omegaR}
\end{equation}
where $\alpha$ and $\beta$ are strengths of, respectively, Rashba and Dresselhaus couplings.
It is convenient to introduce the spin density $s^{\mu}_{\bf k}={\rm Tr}[\sigma_{\mu} \hat{\rho}_{\bf k}]/2$, where $\mu=x,y,z$,
so that eq.(\ref{kieq}) can be rewritten as
\begin{equation}
\dot{{\bf s}}_{\bf k}-e{\bf E}\cdot\nabla_{\bf k}{\bf s}_{\bf k}={\bf s}_{\bf k}\times {\bf H}+{\bf s}_{\bf k}\times {\bf \Omega}_{\bf k} +\sum_{\bf   k'} {\bm J}_{{\bf k' \rightarrow \bf k}},
\label{balance}
\end{equation}
where ${\bm J}_{{\bf k' \rightarrow \bf k}}$ is the stochastic spin current, in the momentum space, due to scattering between states ${\bf k'}$ and ${\bf k}$. For spin conserving scatterings, ${\bm J}_{{\bf k' \rightarrow \bf k}}=-{\bm J}_{{\bf k\rightarrow \bf k'}}$.


Consider a scattering channel that connects states at ${\bf k'}$ and ${\bf k}$. Let $a_{{\bf k}, \ua}, a_{{\bf k},\da}$ be the annihilation operators of an electron with momentum ${\bf k}$ for spin up and down, respectively.  For such scatterings, without spin-flipping, the evolution of spin-up and spin-down electrons during a small time interval $t$ is described by a scattering matrix \cite{blanter-rev}:
 \begin{eqnarray}
 \left( \begin{array}{l}
a_{{\bf k},\ua/\da}(t) \\
a_{{\bf k'}, \ua/\da}(t)
\end{array} \right) = \left(
{\rm \begin{array}{cc}
R_{\bf k k}& T_{\bf k k'} \\
T_{\bf k' k}& R_{\bf k' k'}
\end{array}}
\right)
\left( \begin{array}{l}
a_{{\bf k}, \ua/\da}(0) \\
a_{{\bf k'}, \ua/\da} (0)
\end{array} \right),
\label{scatt}
 \end{eqnarray}
 where $T_{\bf k k'}$ and $R_{\bf k k}$ are, respectively, time-dependent transmission and reflection amplitudes.
 The spin operator is defined as $\hat{\bf s}_{\bf k}=\frac{1}{2}\Psi_{\bf k}^{\dagger}{\bm \sigma}\Psi_{\bf k}$, with $\Psi_{\bf k}=(a_{{\bf k}, \ua}, a_{{\bf k}, \da})^T$. If we assume that scattering is weak, then $|T_{\bf k k'}|^2= |T_{\bf k' k}|^2\ll 1$, $|R_{\bf k k}|^2=1-|T_{\bf k k'}|^2$. The spin current operator  due to such scatterings is defined by:
$\int_0^t dt' {\bm \hat{J}}_{{\bf k' \rightarrow \bf k}} (t') \equiv  \hat{\bf s}_{\bf k}(t)-\hat{\bf s}_{\bf k}(0) $.

 To determine first two cumulants of the spin current  \cite{blanter-rev}, one can take the trace of the first and second powers of $\hat{\bf s}_{\bf k}(t)-\hat{\bf s}_{\bf k}(0)$ with the density matrix ${\bf \hat{\rho}_{\bf k}^{\rm st}}+{\bf \hat{\rho}_{\bf k}}$, where ${\bf \hat{\rho}^{\rm st}}_{\bf k}$ is the density matrix at the steady state and ${\bf \hat{\rho}}_{\bf k}$ is  due to the currently present spin fluctuation \cite{note1}. The steady state density matrix is approximated by a spin-diagonal matrix ${\bf \hat{\rho}^{\rm st}}_{\bf k} \approx f_{\bf k}\hat{1}_{\bf k}$, where $f_{\bf k}=f(\epsilon_{\bf k})$ is the  Dirac-Fermi distribution over energy $\epsilon_{\bf k}$ and $\hat{1}_{\bf k}$ is a unit matrix in the spin space of fermions with momentum ${\bf k}$. This is equivalent to disregarding terms $O(E^2)$ and higher order corrections in $\Omega/\epsilon_F$, where $\epsilon_F$ is the Fermi energy, in the final expression for the spin correlator.
 We should also assume that there are no initial correlations between different phase
 space volumes and between spin currents at different time moments.   Scattering probability, $|T_{\bf k k'}|^2$, is linearly growing with time when energies of ${\bf k'}$ and ${\bf k}$ are the same. Introducing the scattering rate $\omega_{\bf k,k'}=|T_{\bf k k'}|^2/t$, we then find
 \begin{eqnarray}
 && \la {\bm J}_{{\bf k' \rightarrow \bf k}} \ra= \omega_{\bf k,k'} \left({\bf s}_{\bf k'}- {\bf s}_{\bf k}\right),\\
\nonumber &&\la J^{\mu}_{{\bf k \rar \bf k'}}(t') J^{\nu}_{{\bf k_1 \rar \bf k_1'}}(t) \ra -\la J^{\mu}_{{\bf k \rar \bf k'}}(t') \ra \la J^{\nu}_{{\bf k_1 \rar \bf k_1'}}(t) \ra = \\
 &&\omega_{{\bf k  \bf k'}} f_{\bf k}(1-f_{\bf k'}) \delta (t-t')  \delta_{\mu \nu} (\delta_{{\bf k  \bf k_1}} \delta_{{\bf k' \bf k_1'}} -
\delta_{{\bf k \bf k_1'}}\delta_{{\bf k'  \bf k_1}}).
 \label{corr5}
\end{eqnarray}




Let's introduce variables describing coarse-grained spin characteristics:
\begin{eqnarray}
{\bf S}_0=\sum_{{\bf k}}{\bf s}_{\bf k}, \quad {\bf S}_{1, 2}=\sum_{{\bf k}}{\bf s}_{\bf k}k_{x, y},
\label{S12eq}
\end{eqnarray}
and the transport life time $\tau_{\rm tr}$:
\begin{eqnarray}
\frac{1}{\tau_{\rm tr}}=\sum_{{\bf k}'} \omega_{{\bf k}, {\bf k}'} [1-\cos(\varphi-\varphi')],
\end{eqnarray}
with $\varphi$ and $\varphi'$ are angles of, respectively, ${\bf k}$ and ${\bf k}'$ taken at the Fermi surface.
By summing over ${\bf k}$ in Eq.~(\ref{balance})   we then  obtain
\begin{eqnarray}
\dot{\bf S}_0={\bf S}_0\times{\bf H}+{\bf S}_1\times{\bm \Omega}_1+{\bf S}_2\times{\bm \Omega}_2.
\label{s00}
\end{eqnarray}
Due to $k_BT \ll \epsilon_F$,  where $T$ is temperature, we can approximate $ \sum_{\bf k} k_{x,y}^2 {\bf s}_{\bf k} \approx (k_F^2/2) {\bf S}_{0}$, where $k_F$ is the Fermi momentum.
Multiplying Eq.~(\ref{balance}) by $k_x$ and $k_y$ and summing over ${\bf k}$ we then find
\begin{eqnarray}
\dot{\bf S}_1&=&-eE_x{\bf S}_0+\frac{k_F^2}{2}{\bf S}_0\times {\bm \Omega}_1-\frac{{\bf S}_1}{\tau_{\rm tr}} + {\bm \eta}_1, \label{s1eq}\\
\dot{\bf S}_2&=&-eE_y{\bf S}_0+\frac{k_F^2}{2}{\bf S}_0\times {\bm \Omega}_2-\frac{{\bf S}_2}{\tau_{\rm tr}}+ {\bm \eta}_2,
\label{s2eq}
\end{eqnarray}
where we neglect the terms proportional to $\bf{H}$ because $H\ll1/\tau_{\rm tr}$.   The relaxation terms in (\ref{s1eq}) and (\ref{s2eq}) originate from the mean value of the  stochastic spin current, and the spin noise terms are defined as:
\be
{\bm \eta}_{1,2}=\sum_{{\bf k},{\bf k}'} k_{x,y} ( {\bm J}_{{\bf k' \rightarrow \bf k}}- \la {\bm J}_{{\bf k' \rightarrow \bf k}} \ra),
\label{j12}
\ee
with their averages zero and correlations:
\begin{eqnarray}
\la  \eta^{\mu}_i \eta^{\nu}_j\ra 
= k_F^2 \delta_{\mu \nu} \delta_{ij} D k_BT \delta (t-t') /(2 \tau_{\rm tr}),\,\, i,j=1,2, 
\label{ds21}
\end{eqnarray}
where $D$ is the  density of states  per spin in the full observation region. Linear dependence on temperature $T$ appears in (\ref{ds21}) after integration of $f_{\bf k}(1-f_{\bf k})$ over energy, i.e. it can be traced to the Dirac-Fermi statistics of electrons.


Due to short correlation time of spin currents and due to fast relaxation, first harmonics $S_1^{\mu}$ and $S_2^{\mu}$ will change at fast time-scales, at which variables $S_0^{\mu}$  can be considered constant.
This allows us to express first harmonics, e.g. ${\bf S}_1$,  as functions of ${\bf S}_0$:
\be
{\bf S}_1 = -eE_x\tau_{\rm tr}{\bf S}_0  + \frac{k_F^2 \tau_{\rm tr}}{2}  {\bf S}_0\times{\bm \Omega}_1 +{\bm \kappa}_1(t),
\label{s1xn2}
\ee
where ${\bm \kappa}_1$ is the solution of the equation,
$
 {\dot{{\bm \kappa}}}_1 =  -\frac{{\bm \kappa}_1}{\tau_{\rm tr}} + {\bm \eta}_1(t)
$,
which describes a noise with correlators
%
\be
\la {\kappa}_1^{\mu}(t) \ra = 0, \quad \la {\kappa}_1^{\mu}(t) {\kappa}_1^{\nu}(t') \ra =
\frac{k_F^2 D k_BT }{4} e^{-|t-t'|/\tau_{\rm tr}} \delta_{\mu, \nu}.
\label{ss1}
\ee
We also note that the correlator (\ref{ss1}) has very short decay time $\tau_{\rm tr}$ (less than $10$ps in Rashba 2D electron gas), so at time scales of spin relaxation it can be safely approximated by a white noise
\be
 \la {\kappa}_1^{\mu}(t) {\kappa}_1^{\nu}(t') \ra \approx
\frac{k_F^2 D k_BT \tau_{\rm tr}}{2} \delta(t-t')\delta_{\mu, \nu}.
\label{ss1-app}
\ee


By similarly working out equations for ${\bf S}_2$ and substituting results into (\ref{s00}) we  obtain  equations for slowly changing spin density.
In order to simplify notation, we introduce new parameters $ H_{\rm SO} = 2\lambda eE \tau_{\rm tr}$ with $\lambda=\sqrt{\alpha^2+\beta^2}$ and $E=\sqrt{E_x^2+E_y^2}$, and $\frac{1}{\tau_s} = (2\lambda k_F)^2 \tau_{\rm tr} $ as well as the noise variables ${\bm \xi}$:
\be
{\bm \xi}={\bm \kappa}_1\times{\bm \Omega}_1+{\bm \kappa}_2 \times {\bm \Omega}_2.
\label{noise-slow1}
\ee
Note that $\la \xi^{x} \xi^{y} \ra \ne 0$. Equations for dynamics of  $S^{\mu}_0$ are:
\begin{eqnarray}
\nonumber {\dot{S}}_0^z &=& ({\bf S}_0\times{\bf H}_{\rm eff})_z  - \frac{S_0^z}{\tau_s} +\xi_z, \\
  \nonumber  \\
 \label{s0eq2} {\dot{S}}_0^x &=& ({\bf S}_0\times{\bf H}_{\rm eff})_x  - \frac{S_0^x}{2\tau_s}-\frac{S_0^y\sin(2\phi)}{2\tau_s} +\xi_x,\\
   \nonumber   \\
\nonumber  {\dot{S}}_0^y &=&  ({\bf S}_0\times{\bf H}_{\rm eff})_y-\frac{S_0^y}{2\tau_s}-\frac{S_0^x\sin(2\phi)}{2\tau_s} +\xi_y,
\end{eqnarray}
with ${\bf H}_{\rm eff}={\bf H}+{\bf H}_{\rm SO}$, in which
\be
{\bf H}_{\rm SO}=H_{\rm SO}(-\sin(\theta+\phi), \cos(\theta-\phi), 0),
\label{heff}
\ee
where $\theta$ denotes the angle that the in-plane electric field makes with x-axis, and $\tan\phi=\beta/\alpha$. 
 Eqs. (\ref{s0eq2}) and (\ref{heff})
 show that $\tau_s$ is a characteristic relaxation time for the out-of-plane spin component of the fluctuation, and ${\bf H}_{\rm SO}$ is the effective magnetic field which is induced by the  electric current.

By taking Fourier transform of (\ref{s0eq2}), e.g. $S^{z}_0(\omega)=\int_{-T_m/2}^{T_m/2} dt e^{i\omega t} S^z_0(t)$ with $T_m$ being the measurement time, which is  much larger than $\tau_s$,  we obtain the correlator in the frequency domain. When  $|{\bf H}_{\rm eff}| \gg 1/\tau_s$ and a magnetic field is in-plane of the sample, we find that the noise power is given by
\be
P(\omega)\equiv\frac{\la S_0^z (-\omega) S_0^z(\omega) \ra}{T_{m}} = \frac{D k_BT /\tau_e}{(\omega-\omega_L)^2 + 1/\tau_e^2},
\label{ss0}
\ee
where  $\omega_L=|{\bf H}_{\rm eff}|$ and the effective spin relaxation time is $\tau_e^{-1}=\tau_s^{-1}(3-\sin2\varphi \sin2\phi)/4$ in which $\varphi$ is the angle between the direction of ${\bf H}_{\rm eff}$ and the x-axis. Eq. (\ref{ss0}) shows that the shape of the noise power spectrum is Lorentzian but the Larmor frequency is influenced by the electric field and the peak width is  renormalized by a factor, which depends on the direction of the magnetic field and the spin orbit coupling anisotropy.

\begin{figure}
{\includegraphics[width=0.9\columnwidth]{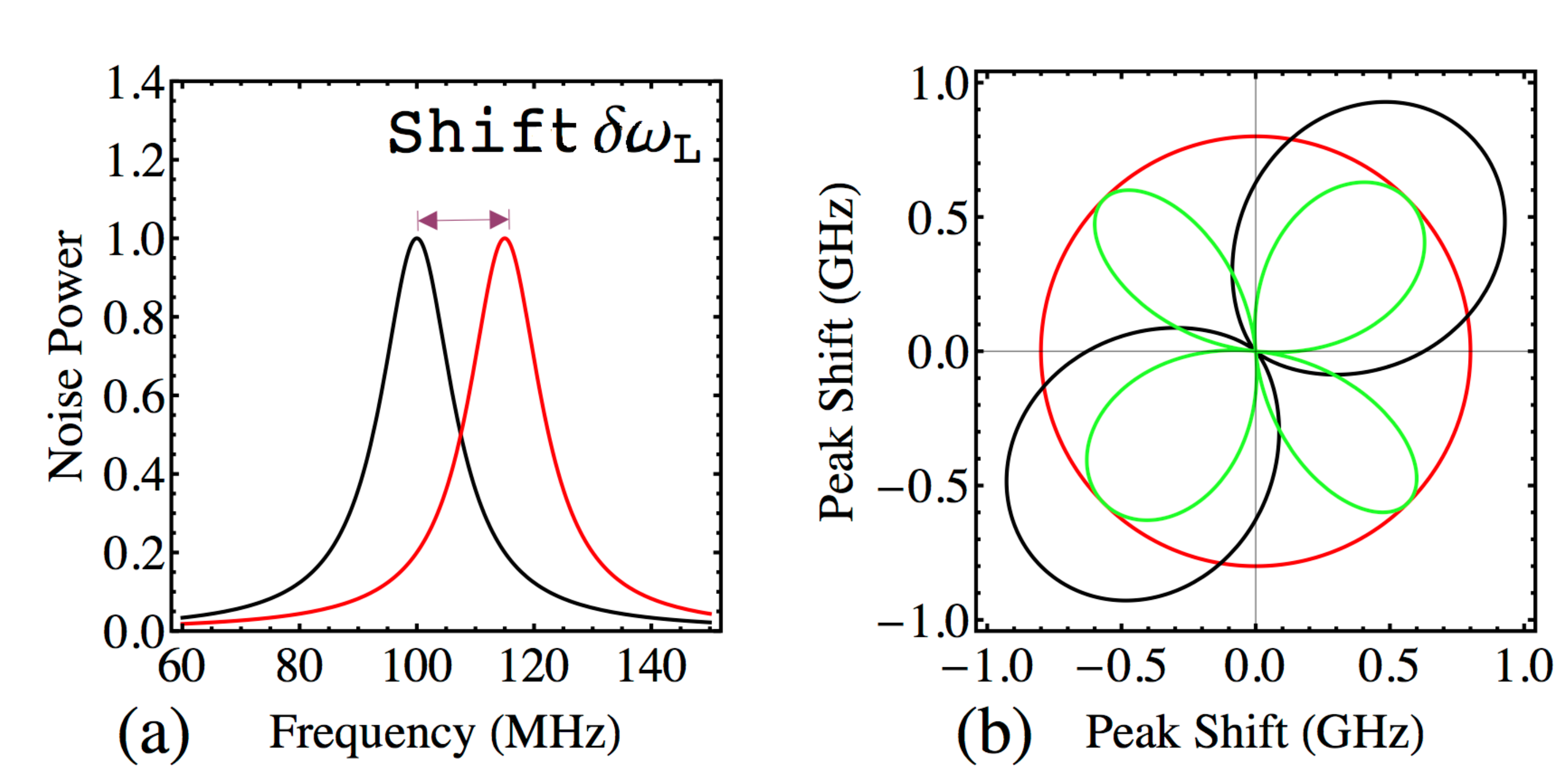}}
\caption{(a) The spin noise power spectrum without and with an in-plane electric field (black and red curves, respectively) in strained bulk GaAs. Maximum of the peak is normalized to 1. The peak is shifted  by an amount of $\delta\omega_L\equiv \omega_L- H$. Here, $H=100$MHz, $H_{\rm SO}=15$MHz, $\tau_s^{-1}=10$MHz. (b) The polar plot of the peak shift $|\delta\omega_L|$ in a $2$D electron gas as a function of $\theta$, i.e. the angle of an in-plane electric field with x-axis. Here $E=12$V/cm, and  the magnetic field is  always perpendicular to the electric field. Red, black and green curves correspond to $\alpha=\lambda\, (\beta=0)$, $\alpha=\beta$ and $\alpha=0\, (\beta=\lambda)$, respectively.} \label{shift}
\end{figure}

{\it Applications.}
A straightforward experimentally testable prediction of Eqs. (\ref{heff}) and (\ref{ss0}) is that the electric field shifts the position of the maximum of the Lorentzian peak by the amount $\delta \omega_L \sim 2eE\lambda \tau_{\rm tr}$,
as we show in Fig.~\ref{shift}(a).  The sign of the shift is proportional to $e$, i.e. depends on the sign of the carriers.


Taking the values from \cite{soc} for Rashba coupling in 2D electron gas  at GaAs/AlGaAs interface,
$
\lambda_R= 1.5 \cdot 10^{-13}eV \cdot m
$, relaxation time
$1/\tau_{\rm tr}=10^{-3}$eV,  
and assuming the electric field $E = 12$V/cm, we find
$H_{\rm SO} \sim 800{\rm MHz}$. 
 which is comparable to the spin relaxation rate in such systems \cite{helix}. 
Alternatively, a linear Rashba-type spin orbit coupling is induced in bulk 3D GaAs samples at imposed strains, i.e. $\alpha \sim \varepsilon \equiv (\varepsilon_{xx}-\varepsilon_{yy})$, where $\varepsilon_{\alpha \beta}$ are components of the strain tensor. The effective spin orbit field, $H_{\rm SO}$, induced by an electric field $E=9$V/cm in a 3D GaAs sample with a strain $\varepsilon =0.015$\% was previously determined to be about $\sim 1$Gauss by a local Hanle measurement approach \cite{crooker-07njp}.  Considering that strains can be increased by an order of magnitude,
the field $E\sim 25$V/cm should produce the shift of the conducting peak by $15$MHz, which would be larger than its width ($\sim 10$MHz) and hence clearly observable (Fig.~\ref{shift}(a)).
The magnitude of the peak-shift effect is sensitive to the anisotropy of the spin orbit coupling, and hence, in a 2D electron gas, depends on the direction of the electric field, as illustrated in Fig.~\ref{shift}(b). In fact,  measuring the shift of the Larmor frequency
at two transverse directions of the external electric field, one can determine strengths of the Rashba and Dresselhaus couplings separately.

\begin{figure}
\scalebox{0.2}[0.2]{\includegraphics{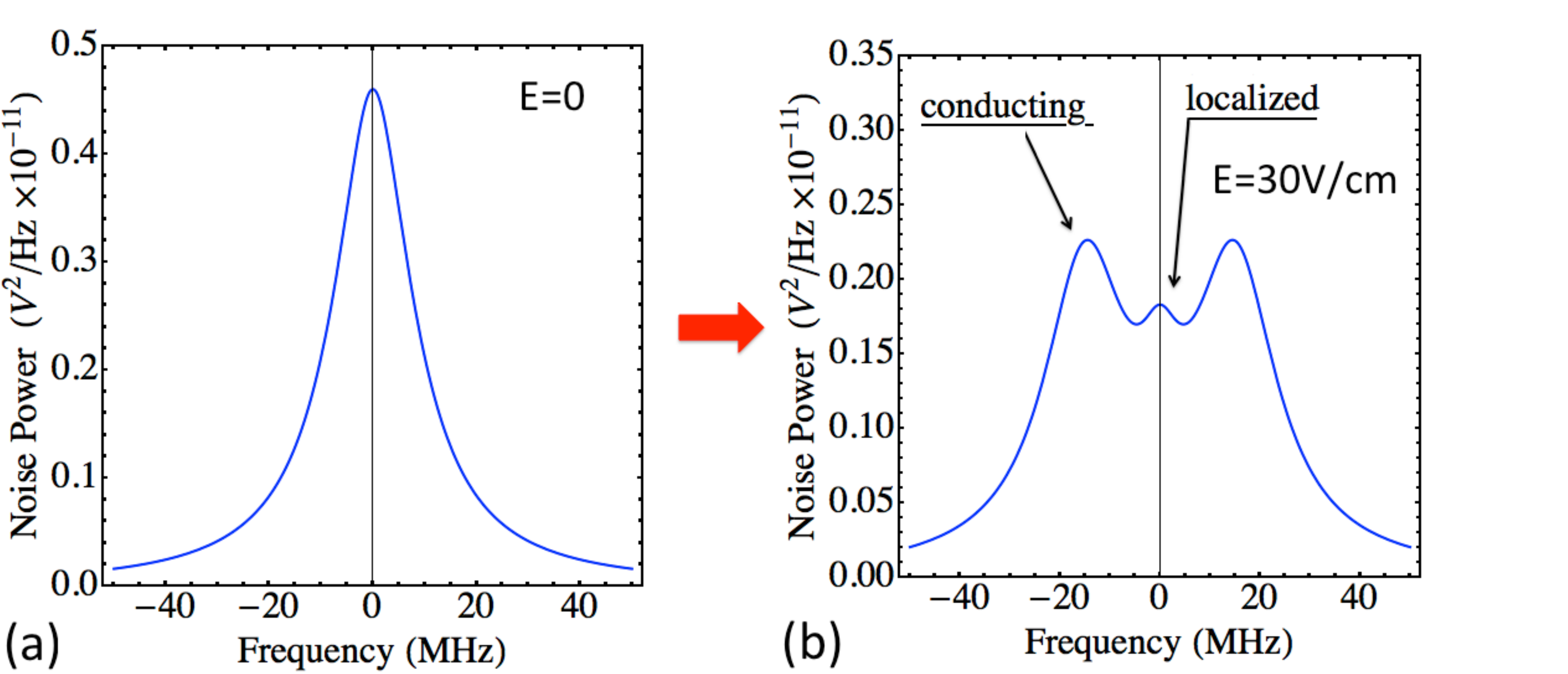}}
\hspace{-2mm}\vspace{-4mm}
\caption{Estimate of the noise power spectrum (a) at zero electric field and (b) at an in-plane electric field $E=30$V/cm for a GaAs sample with comparable numbers of conducting and strongly localized electrons ($n\sim 3\cdot 10^{14}$cm$^{-3}$). Spin relaxation times for conducting electrons is $\tau_s = 100$ns and for localized electrons $\tau_{\rm loc}=500$ns. Electric field $E=30$V/cm at strain $\varepsilon = 0.15$\% splits the peak at zero frequency into two peaks at $\pm H_{\rm SO} \sim 15$MHz due to conducting electrons and a peak that remains at zero frequency, which is produced by localized electrons.
} \label{split}
\end{figure}

Another application of the peak-shift effect can be  in studies of localized states at the presence of conducting electrons.
At low  doping, below the conducting-insulating phase transition  \cite{shklovskii,mott}, there can be donor impurities that are well separated from each other.
 If the distance between impurities exceeds some critical value $R\sim 200$nm, electron hopping between them will be strongly suppressed and localized electron states near such impurities become akin to localized states in
 quantum dots \cite{cond-SNS}.
 Here  we predict that spin noise from conducting and localized electrons can be distinguished due to their different behavior in an applied electric field. Namely, the noise power spectrum for mobile electrons will experience a displacement in the electric field, as explained above,  but the power spectrum of localized electron spins will stay intact, as is shown in Fig.~\ref{split}.

In order to estimate this effect,
 consider a 3D strained GaAs sample with an arbitrary donor concentration $n$. Let $V=4\pi R^3/3$ be the volume that is needed for  a localized electron of a donor impurity  to be well separated from other impurities. The probability that a localized state is not overlapped with any other electron state is given by $e^{-nV}$, so the concentration of impurities whose electron wave functions remain well separated from other donors is given by
\be
n_{\rm loc}(n) = ne^{-nV}.
\label{nloc}
\ee
One can find that for $R=200$nm and a doping above the metal-insulator transition ($n\sim 10^{16}$cm$^{-3}$), the number of such well localized states is negligibly small. However, at lower doping ($n\sim 10^{14}$cm$^{-3}$), the number of well localized and thermally activated conducting electrons, and hence their contribution to the spin noise power, can be made comparable.

The physics of  spin relaxation of isolated localized electrons is expected to be dominated by the hyperfine coupling in essentially the same way as in the spin of electron-doped InGaAs quantum dots, which was discussed in \cite{sinitsyn-12prl}. 
The theory \cite{sinitsyn-12prl} predicts that if the magnetic field is set to zero the localized states of a single donor impurities  produce a sharp noise power peak. At low temperatures (below 7-10K), this peak
has a non-Lorentzian  power-law shape at frequencies below 1MHz  with a broader shoulder, whose width  is determined by the typical strength of the quadrupolar coupling of nuclear spins (Fig.~5b in \cite{sinitsyn-12prl}). For GaAs, the latter is in the order of several megahertz. At moderately large temperatures (7-30 K), phonon mediated mechanisms of localized spin relaxation make this peak shape Lorentzian \cite{Li12}.
Fig.~\ref{split} shows that a reasonably strong electric field is sufficient to shift the peak of conducting electrons and distinguish it from the peak of localized states.



{\it Conclusion.} We predict that measuring the spin noise power spectrum at steady non-equilibrium conditions is a promising research direction with applications to parameter estimation and uncovering new phenomena.
We showed that an electric field leads to a measurable shift of the noise power peak of conducting electrons, which can be used for characterizing the anisotropy of the spin orbit coupling and separating the spin noise of localized states from the spin noise of conducting electrons. Future research directions on the non-equilibrium SNS may include effects of an AC electric field,  spin noise measured from optically polarized electrons, studies of high order fluctuation-dissipation relations \cite{buttiker-ft}, and spin noise in the non-Ohmic regime at strong electric fields \cite{crooker-e10}.


\begin{acknowledgments}
{\it Authors thank S. Crooker, D. Smith, A. Saxena, and Yan Li for useful discussions. This work was funded by DOE under Contract No.\
  DE-AC52-06NA25396.}
\end{acknowledgments}

\end{document}